\documentclass[reprint,pra,aps,showkeys,nofootinbib,onecolumn]{revtex4-2}

\usepackage{setspace}
\usepackage{tocloft}
\usepackage{braket}
\usepackage{array}
\usepackage{hyphenat}
\usepackage[normalem]{ulem}
\usepackage{xcolor}
\usepackage{textcomp,gensymb}
\usepackage{nccmath}
\usepackage{textgreek}
\usepackage{wrapfig}
\usepackage{xcolor} 

\usepackage{amsmath,amssymb}
\usepackage{color}
\usepackage{graphicx}
\usepackage{dcolumn}
\usepackage{bm}
\usepackage{cancel}
\usepackage{mathtools}

\usepackage[colorinlistoftodos]{todonotes}
\DeclareMathSymbol{\R}{\mathalpha}{AMSb}{"52}

\usepackage{fancyhdr}
\usepackage[left=2cm, right=2cm, top=2cm]{geometry}

\begin{document}

\title{Free-space Optical Diffraction and the Fisher Information}

\author{Jonathan M. Nichols $^{1,\dagger}$ and Frank Bucholtz $^{2}$*}

\affiliation{%
$^{1}$Optical Sciences Division, Code 5601, Naval Research Laboratory, Washington, DC 20375;\\
$^{2}$Jacobs Technology, Inc., Herndon, VA 20171 ; frankbucholtz2012@comcast.net}

\begin{abstract}
Using the transport-of-intensity approach for free-space optical propagation in the paraxial regime, we show that diffraction is fundamentally related to the Fisher information associated with models of the transverse intensity distribution. By interpreting intensity as a probability density, we show that a) free-space diffraction will always act to flatten the intensity distribution as the beam propagates, and consequently b) diffraction will monotonically minimize the Fisher Information with respect to any parameterization of the intensity distribution model that depends on propagation distance.
\end{abstract}


\maketitle


\section{Introduction}

Diffraction is a well-known phenomenon in optical wave propagation wherein the transverse intensity distribution of a beam changes as the beam propagates. In an optics curriculum, for example, diffraction is typically introduced in the context of an infinite plane wave encountering an aperture or an obstruction. Of course, diffraction also occurs in free space and can be calculated using well-known diffraction integrals \citep{Goodman:17,Saleh:91}. Under certain physically-reasonable assumptions -- the paraxial assumptions -- these diffraction integrals are exact solutions of the paraxial wave equation. The physical interpretation typically assigned to these integrals is that each point on the source aperture itself acts as a source of a spherical wave, and the downrange electric field is the summation of these spherical waves. However, as we will show below, it is not the aperture {\emph{per se}} that causes diffraction but rather the shape of the resulting intensity profile that fundamentally determines beam spreading upon propagation.

There exists an alternate method for modeling free-space diffraction referred to generically as the "transport of intensity" (TOI) approach \citep{Teague:82,Streibl:84,Zuo:20}. Working within the TOI approach, we will show that diffraction will always result in a flattening of the beam's intensity profile as the beam propagates.  As a consequence, free-space diffraction always occurs in such a way as to minimize the Fisher information (FI) associated with a given parametric model of the beam's intensity.

The connection between FI and diffraction was  hypothesized over 35 years ago by Frieden \cite{Frieden:89} where he used a series of logical arguments to conclude  ``In particular, diffraction is seen to be nature's attempt to maximally smear out spatially the probability law on position for each photon forming a diffraction pattern''.  In this paper, we show that the same conclusion arises naturally as a consequence of the TOI approach and we present several relevant examples demonstrating this interpretation of diffraction as a FI minimizer.    

\section{Physics of Optical Beam Propagation \label{sec:physics}}
The propagation of a uniformly-polarized optical beam is a well-known phenomenon in classical optics. The beam can be defined completely at time $t$ and spatial location $(x,y,z)$ in terms of the components of the associated oscillating electric and magnetic field vectors, denoted ${\bf{E}}$ and ${\bf{H}}$, respectively. In general, these time- and space-varying vectors must satisfy Maxwell's equations and a derived set of wave equations -- one for each of the vector components of ${\bf{E}}$ and ${\bf{H}}$.  In many cases, assumptions can be made that simplify the problem. Four important such assumptions are
\begin{enumerate}
\item The beam is paraxial meaning that, over the propagation region of interest, the wavevectors comprising the beam do not stray too far from the average direction of propagation, taken here to be the Cartesian $z-$axis.
\item The beam is monochromatic, comprising field oscillations at a single wavelength $\lambda$ and the sign convention for phase propagation is here taken to be 
$\exp{i(kz-\omega t)}$.
\item The beam is uniformly polarized (a point we return to later in the discussion) since polarization inhomogeneities can create dynamics not captured by the scalar approach in Eq. (\ref{eq:PWE}) below \cite{Nichols:22,Nichols:24}.
\item The propagation is lossless.
\end{enumerate}
With these assumptions, the full wave equation simplifies to an equation known as the scalar paraxial wave equation (PWE) given by 
\begin{align}
\nabla_T^2A+2ik\dfrac{\partial{A}}{\partial{z}}=0
\label{eq:PWE}
\end{align}
where $k=2\pi/\lambda$ is the wavenumber, $\nabla_T^2=\partial_{xx}+\partial_{yy}$ is the transverse part of the Laplacian operator, and $A(x,y,z)$ is the complex envelope of any one of components of the fields ${\bf{E}}$ or ${\bf{H}}$. The total complex amplitude of the paraxial field component is then given by
\begin{align}
U(x,y,z)=A(x,y,z)\exp{(ikz)},
\end{align} that is, a plane wave propagating along $z$ but modulated in space by $A(x,y,z)$ \citep{Saleh:91}.

Starting with the paraxial wave equation (\ref{eq:PWE}), two approaches can be employed to determine the properties of the field at downrange location $z$ given that we have complete knowledge of the field at some prior location $z_1<z$. By far the most popular approach is the use of diffraction integrals (DIs)  \citep{Goodman:17,Saleh:91}.  Given the complex amplitude at the source aperture $A(x_1,y_1,0)$, the solution to (\ref{eq:PWE}) is given by
\begin{align}\label{eq:HFIntegral}
    A(x,y,z)&
    =\left(\frac{k}{2\pi i z}\right)
    \int\int A(x_1,y_1,0)\exp{\left(\dfrac{-i\,k\left((x-x_1)^2+(y-y_1)^2\right)}{2z}\right)}dx_1dy_1
\end{align}
where $(x,y,z)$ are coordinates in the observation plane at downrange distance $z$ and $(x_1,y_1,0)$ are transverse coordinates in the source $(z=0)$ plane. While the solution is exact, it provides little intuition concerning the impact of diffraction on a beam's propagating intensity distribution, $\rho(x,y,z)\equiv A(x,y,z)A^*(x,y,z)$.  While the form of (\ref{eq:HFIntegral}) suggests a ``spreading'' of the complex amplitude, the topic of interest in this paper is the behavior of the local structure of the intensity as the beam propagates as a function of the parameters governing its evolution. We will nonetheless find use for DIs when calculating beam amplitude distributions for the examples in Section~\ref{sec:Examples}. 

As an alternative to the DI approach, we employ a less well-known approach referred to in the literature as transport-of-intensity (TOI)  \citep{Teague:82,Streibl:84,Zuo:20}. The approach reformulates the diffraction problem as a coupled set of transport equations describing the evolution of optical intensity and phase.  As we will see, these equations naturally reveal the ``information-theoretic'' nature of free-space diffraction, a connection that is completely hidden in the DI approach.  Moreover, the TOI equations can be used to connect directly the process of diffraction with the minimization of Fisher Information.

To proceed with the TOI approach we assume that $A$ can be written in polar form as 
\begin{align} \label{eq:TOIForm}
A(x,y,z)=A_0\rho^{1/2}(x,y,z)e^{i\phi(x,y,z)}
\end{align}
where the real constant $A_0$ has units of electric field, $\phi(x,y,z)$ is the overall phase distribution of the beam, and $\rho(x,y,z)$ is a non-negative, dimensionless function that describes the spatial intensity distribution of the beam.
Substituting (\ref{eq:TOIForm}) into the PWE (\ref{eq:PWE}) yields two real-valued equations
\begin{subequations}\label{eq:TOI}
\begin{align}
&\dfrac{\partial\rho}{\partial z}+\nabla_T\cdot\left(\rho \vec{v}\right)=0, \label{eq:ContIntensity}\\[8pt]
&\dfrac{D\vec{v}}{Dz}=\dfrac{-1}{\rho}\nabla_T\cdot{\bf{P}}\equiv \vec{a}_T,
\label{eq:ContMomentum}
\end{align}
\label{eqn:transport}
\end{subequations}
obtained by setting (respectively) the resulting imaginary and real parts of (\ref{eq:PWE}) equal to zero and taking the transverse gradient of the latter.  Here $D(\cdot)/Dz$ denotes the total, or material, derivative commonly used in fluid mechanics (see e.g. \cite{Panton:96}) and is defined by
\begin{align}\label{eqn:TotalvDer}
\dfrac{D\vec{v}}{Dz}\equiv\dfrac{\partial \vec{v}}{\partial z}+(\vec{v}\cdot\nabla_T) \vec{v},
\end{align}
where $\nabla_T$ is the transverse Cartesian del operator $\nabla_T=\partial_x\hat{x}+\partial_y\hat{y}$.
Equation (\ref{eq:ContIntensity}) expresses the conservation of intensity.  That is, while the transverse location of each local parcel of intensity can change due to the transverse phase gradient $\vec{v}\equiv k^{-1}\nabla_T\phi$, Eq.~(\ref{eq:ContIntensity}) states that the total intensity integrated over the transverse plane is conserved.

The vector field
$\vec{v}$ is effectively a dimensionless "velocity" that defines the movement in the transverse plane of an infinitesimal volume of intensity at position $(x,y,z)$ as the beam propagates forward.  Specifically, $\vec{v}$ is the change in optical path in the transverse direction per unit change in the direction of propagation \cite{Nichols:23}. $D\vec{v}/Dz$ is therefore appropriately viewed as a transverse acceleration $\vec{a}_T$. 

This acceleration is governed by the dimensionless tensor 
\begin{align}\label{eq:Pdef}
{\bf{P}}(x,y,z)&=\dfrac{-\rho(x,y,z)}{4k^2}\left[\begin{array}{cc} \dfrac{\partial^2}{\partial x^2} & \dfrac{\partial^2}{\partial x\partial y}\\[12pt]
\dfrac{\partial^2}{\partial x\partial y} & \dfrac{\partial^2 }{\partial y^2}
\end{array}
\right]\ln\left(\rho(x,y,z)\right)
\end{align}
which we refer to as the "optical pressure tensor" since it plays a role analogous to physical pressure in fluid mechanics.\footnote{However, a crucial difference is that the optical pressure acts only transversely to beam propagation and has no component along the direction of propagation.}   It is then the divergence of this pressure,
\begin{align}
\nabla_T\cdot{\bf{P}}\equiv&\Big(\partial_xP_{1,1}+\partial_yP_{1,2}\Big)\hat{x}\nonumber\\
&\qquad+\Big(\partial_xP_{2,1}+\partial_yP_{2,2}\Big)\hat{y},
\end{align}
that dictates the optical path taken by each local region of intensity.  Here $P_{i,j}$ denotes the $(i,j)$ element of the pressure tensor ${\bf{P}}$.

We note that ${\bf P}(x,y,z)$ represents an {\it internal} pressure as it arises due to the expected transverse spatial distribution of photons that comprise the beam as opposed to being externally applied.  The ``internal'' nature of this term has also been noted in similar continuum mechanics models (see e.g., \cite{Ghosh:83,Yahalom:18}). 
Interestingly, in Yahalom \cite{Yahalom:18} it was conjectured that this pressure term ``...can be interpreted as Fisher information ...'', a statement we put on a more rigorous footing in what follows.  

\section{The Fisher Information for Optical Intensity}
\subsection*{Probability Density and Conditional Probability Density}
Let $W$ denote a continuous random variable corresponding to some physical observable and let $w$ denote a possible outcome of a measurement of $W$. Let $p_W(w)$ denote the probability density function for $W$. The probability density is normalized over the set of all possible outcomes
\begin{align}
\int_{W}p_W(w)dw=1.
\end{align}
  In any measurement of $W$, the probability of observing a value in the closed range $[w_a,w_b]$ is \begin{align}
\mathrm{Pr}\{w\in [w_a,w_b]\}=\int_{w_a}^{w_b}p_W(w)dw.
\end{align}
The concept of probability density can be extended to the case where $W$ is a vector of random variables $\vec{W}=[W_1,W_2,\cdots,W_M]$ with possible outcomes $\vec{w}\equiv \{w_1, w_2, \cdots,w_M\}$. Then
\begin{align}
&\mathrm{Pr}\{w_1\in [w_{1a},w_{1b}], w_2\in [w_{2a},w_{2b}], \cdots ,w_M\in [w_{Ma},w_{Mb}]\}\nonumber
\\
&\qquad\qquad\qquad\qquad\qquad=\int_{w_{1a}}^{w_{2a}}\int_{w_{2a}}^{w_{2b}}\cdots \int_{w_{Ma}}^{w_{Mb}}p_{\vec{W}}(\vec{w})dw_1dw_2\,....\,dw_M.
\end{align}

The probability density function is thus a model for the uncertainty when measuring physical phenomenon and, as with any model, it can be parameterized. Let $\Theta=\{\theta_1, \theta_2, \cdots, \theta_D \}$ be an array of $D$ parameters that characterize the model. Then $p_{\vec{W}}(\vec{w}|\Theta)$ denotes the conditional probability density under the assumption that the model is characterized by the specific parameter values $\Theta$.

\subsection{Optical Intensity as a Probability Model for Photons}

Viewing diffraction in the context of Fisher information requires interpreting the intensity distribution as a probability model for photon position -- the same interpretation taken by Frieden \cite{Frieden:89} in relating optical intensity to Fisher information.  

In optics, intensity $I(x,y,z)$ is a measure of the time-averaged optical energy per unit time passing through a small area element $\Delta x\Delta y$ perpendicular to the propagation direction, centered on some location $(x,y,z)$ in a transverse plane at distance $z$ from the source, and    
\begin{align}
I(x,y,z)\propto A_0^2\,\rho(x,y,z).
\end{align}
Equivalently, intensity $I(x,y,z)$ is the average number of photons that pass through the same area element per unit time multiplied by the energy $hc/\lambda$ of each photon.

To obtain Fisher information we focus on the shape of the intensity distribution and how that distribution shape evolves during propagation. Subsequent calculations and the interpretation of results will be greatly facilitated if we scale all distances by the optical wavelength $\lambda$,
\begin{align*}
\rho(x,y,z)\longrightarrow\rho\left(\frac{x}{\lambda}, \frac{y}{\lambda},\frac{z}{\lambda}
\right).
\end{align*}
We can further always introduce a multiplicative constant to $\rho$ so that probability normalization is satisfied, that is,
\begin{align}\label{eq:RhoNormalization}
\int_{-\infty}^{\infty}\int_{-\infty}^{\infty}
\rho\left(\frac{x}{\lambda}, \frac{y}{\lambda},\frac{z}{\lambda}
\right)d\left(\frac{x}{\lambda}\right)d\left(\frac{y}{\lambda}\right)=1.
\end{align}
Going forward we will use the following notation for scaled quantities, 
\begin{align}\label{eq:ScaledQuantities}
\begin{matrix}
\tilde{x}=x/\lambda, & \quad \tilde{y}=y/\lambda, & \quad \tilde{z}=z/\lambda,
\end{matrix}
\end{align}
where the tilde indicates that the quantity is scaled by the optical wavelength. We note that any quantity that depends on position or distance must be scaled properly as well. Thus parameters corresponding to length are dimensionless or, equivalently, have dimensions $m/m$.
Then $\rho{(\tilde{x},\tilde{y},\tilde{z})}$ is the probability density corresponding to optical intensity $I(x,y,z)$.  And $\rho{(\tilde{x},\tilde{y}|\,\tilde{\Theta}_z)}$ denotes a conditional probability density where $\widetilde{\Theta}_z=\{\tilde{\theta}_{1z},\tilde{\theta}_{2z}, \cdots\}$ is an array of model parameters and where each parameter itself may be parameterized by $\tilde{z}$. This notation allows for the fact that the shape of the intensity distribution, as determined by $\widetilde{\Theta}_z$, can change as a function of propagation distance. 
Hence, for given values of $\tilde{z}$ and model parameters $\widetilde{\Theta}_z$, 
\begin{align}
\int_{-\Delta \tilde{x}/2}^{\Delta \tilde{x}/2}\int_{-\Delta\tilde{y}/2}^{\Delta \tilde{y}/2}\rho(\tilde{x},\tilde{y}|\widetilde{\Theta}_z)d\tilde{x}d\tilde{y}
\end{align}
is the probability that a photon from the source will pass the transverse plane at normalized propagation distance $\tilde{z}$ through an area of size $\{\Delta \tilde{x}, \Delta \tilde{y}\}$ centered on scaled transverse position $(\tilde{x},\tilde{y})$.
\\

\subsection{Fisher Information}

The Fisher information (FI) is commonly used in estimation theory to quantify the amount of information about model parameters $\widetilde{\theta}_{iz},~i=1\cdots D$  that can be obtained from observed data $\rho(\tilde{x},\tilde{y})$.  When applied to the conditional probability density for optical intensity, the symmetric, $D\times D$ Fisher Information Matrix (FIM) becomes \cite{Whalen:95}

\begin{align}\label{eqn:FIMDefinition}
    {\mathcal{F}}_{ij}(\widetilde{\Theta}_z)&=-\int\int_{\mathbb{R}^2}\rho(\tilde{x},\tilde{y}|{\widetilde{\Theta}_z})\frac{\partial^2\left(\ln\left[\rho(\tilde{x},\tilde{y}\,|\,{\widetilde{\Theta}})\right]\right)}{\partial \theta_i\partial \theta_j}  
    d\tilde{x}d\tilde{y},\quad i,j\in [1,D].
\end{align}

Alternatively, the FIM can be expressed as an expected value,
\begin{align}
{\mathcal{F}}(\widetilde{\Theta}_z)&=-\mathbb{E}\left[{\bf H}_{\widetilde{\Theta}_z}\big(\ln\big(\rho(\tilde{x},\tilde{y}\,|\,\widetilde{\Theta}_z\big)\big)\right],
\label{eq:FIM2}
\end{align}
where the Hessian matrix ${\bf{H}}_{\widetilde{\Theta}_z}(f)$ of a function $f$ of parameters $\widetilde{\Theta}_z$ is defined by
\begin{align}
{\bf{H}}_{\widetilde{\Theta}_z}(f)=\left[\dfrac{\partial^2f(\vec{x})}{\partial \tilde{\theta}_{iz}\partial \tilde{\theta}_{jz}}\right],\quad i,j\in [1,D].
\label{eq:Hessian}
\end{align}

A useful single-valued representation of the Fisher matrix is its determinant 
\begin{align}
{\bf{F}}(\widetilde{\Theta}_z)=\mathrm{det}\left(\mathcal{F}(\widetilde{\Theta}_z)\right),
\end{align}
often referred to as, simply,
"the Fisher information" (FI). Thus the FI associated with the photon probability model $\rho(\tilde{x},\tilde{y}\,|\,\widetilde{\Theta}_z)$ is the determinant of (\ref{eqn:FIMDefinition}).

We will also be interested in the rate of change of the FI as the beam propagates denoted
\begin{align}\partial_{\tilde{z}}{\bf F}(\widetilde{\Theta}_z)\equiv \partial{\bf F}(\widetilde{\Theta}_z)/\partial {\tilde{z}}.
\end{align}

Next we show that the dynamics governed by (\ref{eqn:transport}) will always act to flatten the distribution.  As the intensity distribution flattens or "smears" with propagation, the FI associated with this distribution must likewise approach zero.

\section{Diffraction as a Minimizer of Fisher Information \label{sec:minFIM}}

Given the physics described in Section (\ref{sec:physics}) and the definition of the FI of the previous section, we are now in a position to show how the former minimizes the latter.
We begin by noting that Eqs. (\ref{eqn:transport}) can be conveniently re-written in Lagrangian coordinates, $x,y,z\rightarrow x_z,y_z$, whereby the locations of each local region of intensity $\rho(x_z,y_z)$ are now parameterized by the propagation distance $z$ \cite{Price:06}.  Each parcel of intensity at $z=0$ can be thought of as ``following'' the trajectories $(x_z,y_z)$.  These coordinates are themselves dependent variables that evolve with $z$.  To see this, note that in Lagrangian coordinates $\vec{v}\equiv d\vec{x}_z/dz$ and Eqn. (\ref{eq:ContMomentum}) becomes (see \cite{Nichols:22},\cite{Nichols:23} for details),
\begin{align}
    \frac{d^2 \vec{x}_z}{dz^2}&=-\frac{1}{\rho}\nabla_{\vec{X}_z}\cdot {\bf P}\,=\vec{a}_T.
    \label{eqn:momLagrange}
\end{align}

This means that the path taken by each local region of intensity is governed by the negative of the transverse divergence of ${\bf P}$. To explore this term further, re-write the diffractive tensor ${\bf P}$ as
\begin{align}
    {\bf P}&=\frac{\nabla_{\vec{X}_z}\rho(x_z, y_z)\otimes\nabla_{\vec{X}_z}\rho(x_z, y_z)}{\rho(x_z,y_z)}-{\bf H}_{\vec{x}_z}(\rho(x_z, y_z))
    \label{eqn:stationary}
\end{align}
which comprises two terms: a) the outer-product of transverse gradients of the intensity, and b) the Hessian of the intensity.  Note that the Hessian appearing in Eq. (\ref{eqn:stationary}) is {\it not} the same as the Hessian (\ref{eq:Hessian}) as it is defined with respect to the Lagrangian spatial coordinates $\vec{X}\equiv\{x_z,y_z\}$ as opposed to the parameter values $\widetilde{\Theta}_z$.

If we restrict attention for the moment to stationary points $(x_{zs}, y_{zs})$ on the intensity distribution, that is, either peaks or troughs, the  gradients vanish and the pressure tensor can be expressed locally, at an arbitrary $z$, as 
\begin{align}
    {\bf P}=-{\bf H}_{\vec{x}_z}(\rho(x_{zs}, y_{zy})).
\end{align}
According to Eq. (\ref{eqn:momLagrange}), the local acceleration of intensity at these stationary points is governed only by the divergence of the Hessian of the intensity.  The resulting acceleration vector will always point away from a local intensity maximum and the intensity always moves accordingly, that is, $\vec{x}_z$ carries intensity away from maxima.  The end result of this ``local'' spreading of intensity near stationary points must be a flattening of the overall distribution.  Because the propagation is assumed to be lossless (see again Eq. \ref{eq:ContIntensity}) the total power is conserved and "spreading" is equivalent to "flattening." Thus, we can state that {\it The physics of diffraction will monotonically flatten the intensity distribution as the beam propagates and monotonically minimize the Fisher information}.  This is the key piece of insight surrounding the problem physics afforded by the TOI model (\ref{eqn:transport}) and is valid independent of whether normalized or unnormalized lengths are used.

No matter how the intensity distribution is parameterized, the ultimate outcome of diffraction will, for large enough $\tilde{z}$, be a flattened distribution but one that must go to zero as $\tilde{z}\rightarrow \pm\infty$. This must be true since  $\rho(\tilde{x},\tilde{y})$ must remain integrable and positive definite for all $\tilde{z}$ values (Eq.~\ref{eq:RhoNormalization}). Thus, the physics of diffraction will act to drive each entry of the FIM (\ref{eqn:FIMDefinition}) toward zero as $\tilde{z}\rightarrow \infty$ and,  likewise, ${\bf F}(\tilde{\theta}_z)\rightarrow 0$ as $\tilde{z}\rightarrow\infty$.  Alternatively, a histogram estimator of intensity could be employed in which case the parameters would be the intensity values in each ``bin'' of discretized intensity.  As the number of parameters becomes large (bin size shrinks) ${\bf H}_{\Theta_z}\rightarrow {\bf H}_{\vec{X}_z}$ in which case the minimization of the corresponding FI is the direct result of (\ref{eqn:momLagrange}). This is certainly a sensible result.  As a beam loses structure in its intensity distribution, the measured intensity carries less and less information with which to estimate its model parameters.

To conclude this section, Table \ref{table:1} provides a listing of the key expressions used in deriving the FI as a function of parameterized intensity distribution $\rho(\tilde{x},\tilde{y}|\,\widetilde{\Theta}_z)$. With the scaled length parameters used in the analyses in this paper, \emph{all} the relevant derived quantities ${\bf{H}}$, ${\bf{P}}$, ${\bf{a}}_T$, ${\mathcal{F}}_{ij}$, ${\bf{F}}$ and $\partial_{\tilde{z}}{\bf{F}}$ are dimensionless. This may not be true in general.

\begin{table}[h!]\label{table:Nomenclature}
\centering
\begin{tabular}{ |p{3.5cm}|p{7.5cm}|  }
 \hline
 \multicolumn{2}{|c|}{ Derived Quantities} \\
 \hline\hline
 Hessian Matrix &${\bf H}(\tilde{x},\tilde{y}\,|\,\widetilde{\Theta}_z)=\dfrac{\partial^2 \ln\left(\rho(\tilde{x},\tilde{y}|\,\widetilde{\Theta}_z)\right)}{\partial \tilde{\theta}_{iz}\partial \tilde{\theta}_{jz}}$ \\*[0.25in]
 Pressure Tensor   & ${\bf{P}}(\tilde{x},\tilde{y})=\dfrac{-\rho(\tilde{x},\tilde{y})}{4(2\pi)^2}\left[\begin{array}{cc} \dfrac{\partial^2 }{\partial \tilde{x}^2} & \dfrac{\partial^2 }{\partial \tilde{x}\partial \tilde{y}}\\
\dfrac{\partial^2 }{\partial \tilde{y}\partial \tilde{x}} & \dfrac{\partial^2 }{\partial \tilde{y}^2}
\end{array}
\right]\ln\left(\rho(\tilde{x},\tilde{y}\right))$    \\*[0.35in]
 Acceleration Vector& $\vec{a}_T(\tilde{x},\tilde{y})=\dfrac{D\vec{v}}{D\tilde{z}}=\dfrac{-1}{\rho}\tilde{\nabla}_T\cdot{\bf{P}}$ \\*[0.25in]
 Fisher Matrix & ${\mathcal{F}}_{ij}(\widetilde{\Theta}_z)=-\int\int_{\mathtt{R}^2}\rho(\tilde{x},\tilde{y}|\,\widetilde{\Theta}_z){\bf H}(\tilde{x},\tilde{y}|\,\widetilde{\Theta}_z)d\tilde{x}d\tilde{y}$\\*[0.25in]
 Fisher Information &${\bf{F}}(\widetilde{\Theta}_z)=\det\left(\mathcal{F}_{ij}(\widetilde{\Theta}_z)\right)$\\*[0.25in]
 Rate of Change of {\bf{F}}& $\partial_{\tilde{z}}{\bf{F}}(\widetilde{\Theta}_z)=\partial{\bf{F}}(\widetilde{\Theta}_z)/\partial \tilde{z} $ \\*[0.1in]
 \hline
\end{tabular}
\caption{Definitions of the relevant derived quantities. Recall that a tilde indicates that the quantity has been scaled by optical wavelength, the $\tilde{\theta}_{iz}$ are $z-$dependent model parameters, $\widetilde{\Theta}_z=\{\tilde{\theta}_{1z},\tilde{\theta}_{2z}, \cdots\}$ denotes an array of $z-$dependent model parameters, and $\vec{v}=k^{-1}\nabla_T\phi$ is the transverse "velocity" of intensity arising from the transverse phase gradient.
}
\label{table:1}
\end{table}
Importantly, note that the pressure tensor ${\bf{P}}$ and the transverse acceleration vector $\vec{a}_T$ are not part of any estimation process. Instead, they are \emph{physical} characteristics of the diffraction associated with a given intensity distribution $\rho(\tilde{x},\tilde{y})$. Hence, when defining these two quantities, the probability $\rho(\tilde{x},\tilde{y})$ is used -- not the conditional probability. 

In the next Section we illustrate these results with several examples.  

\newpage

\section{Examples}\label{sec:Examples}
\begin{figure}[tbh]
 \centerline{
  \begin{tabular}{c}
  \includegraphics[width=9cm,angle=0]{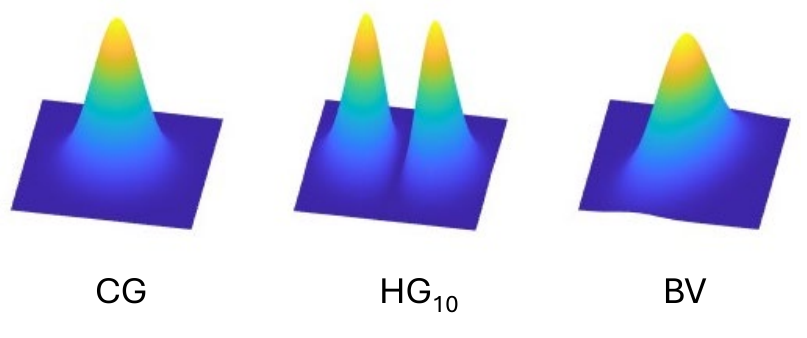}   
  \end{tabular}
  }
   \caption{The three beam examples studied in this Section. CG = circularly-symmetric Gaussian beam, HG$_{10}$ = the $\{1,0\}-$Hermite-Gaussian beam, BV = bivariate Gaussian beam. }
   \label{fig:BeamSummary2}   
\end{figure}

In this section we analyze three specific examples of Gaussian-based optical beams with various intensity distributions, each of which satisfy the PWE. These are i) a circularly-symmetric Gaussian beam (CG), ii) a Hermite-Gaussian beam, in particular the HG$_{10}$ beam, and iii) a bivariate Gaussian beam (BG). The general shape of each distribution is shown in Fig.~\ref{fig:BeamSummary2}. At the end of the Section we discuss briefly the plane wave and the paraboloidal beam .  In each example, we will see that under appropriate $\tilde{z}$-dependent parameterization, the intensity distributions will always evolve in such a way as to monotonically minimize the associated FI.

\subsection{Uniformly-polarized, Circularly-symmetric Gaussian Beam} \label{sec:CCGB}
Consider propagation of a uniformly-polarized, circularly-symmetric Gaussian beam with wavenumber $k=2\pi/\lambda$ corresponding to wavelength $\lambda$ and scaled $\tilde{z}-$dependent standard deviation $\tilde{\sigma}_z$ . The dimensionless probability density associated with the propagating beam intensity is \citep{Saleh:91}
\begin{align} \label{eq:Rho00}
\rho_{00}(\tilde{x},\tilde{y},\tilde{z}|\,\tilde{\sigma}_{z})&=\dfrac{2}{\pi\tilde{\sigma}_z^2}\exp{\left(-2(\tilde{x}^2+\tilde{y}^2)/{\tilde{\sigma}}_z^{2}\right)}.
\end{align}
This conditional probability density, containing the single dimensionless model parameter $\tilde{\sigma}_z$, is obtained as follows.
\begin{enumerate}
\item The scalar electric field distribution in the source plane,  $A_{00}(x_1,y_1,0)$, is assumed to have constant transverse phase distribution and amplitude distribution given by the square root of the source intensity distribution, that is,
\begin{align}
A_{00}(\tilde{x}_1,\tilde{y}_1,0|\,\tilde{\sigma}_0)=\sqrt{\rho_{00}(\tilde{x}_1,\tilde{y}_1,0|\,\tilde{\sigma}_0)}=\Bigg(\dfrac{2}{\pi \tilde{\sigma}_0^2}\Bigg)^{1/2}\exp{\Bigg(\dfrac{-(\tilde{x}^2_1+\tilde{y}_2^2)}{\tilde{\sigma}_0^2}}\Bigg)
\end{align}
where $\tilde{\sigma}_0^2$ is the scaled, dimensionless beam variance at $\tilde{z}=0$.\\
\item The source field $A_{00}(\tilde{x}_1,\tilde{y}_1,0|\,\tilde{\sigma}_0)$ is then propagated forward along the positive $z-$axis using the Huygens-Fresnel (HF) integral (\ref{eq:HFIntegral}) to obtain the transverse field distribution $A_{00}(\tilde{x},\tilde{y},\tilde{z}|\,\tilde{\sigma}_z)$ at downrange distance $\tilde{z}>0$,
\begin{align}
A_{00}(\tilde{x},\tilde{y},\tilde{z}|\,\tilde{\sigma}_z)=
\dfrac{\sqrt{2\pi}\,\,\tilde{\sigma}_0}{\pi\tilde{\sigma}_0^2-i\,\tilde{z}}\exp{\Bigg(\dfrac{-(x^2+y^2)}{\pi\tilde{\sigma}_0^2-i\,\tilde{z}}\Bigg)}.
\end{align}
\item The intensity distribution $\rho_{00}(\tilde{x},\tilde{y},\tilde{z}|\,\tilde{\sigma}_{z})$ for $\tilde{z}>0$ is then obtained simply as the modulus squared of the field distribution
\begin{align}
\rho_{00}(\tilde{x},\tilde{y},\tilde{z}|\,\tilde{\sigma}_{z})=|A(\tilde{x},\tilde{y},\tilde{z}|\tilde{\sigma}_z)|^2
\end{align}
which yields (\ref{eq:Rho00}).
\end{enumerate}

Model parameter $\tilde{\sigma}_z$ is related to optical quantities through

\begin{align}\label{eq:sigz}
\tilde{\sigma}_z
=\left(\dfrac{\tilde{z}^2}{\pi^2\tilde{\sigma}_0^2}+\tilde{\sigma}_0^2
\right)^{1/2}
= \frac{1}{\lambda}\dfrac{(z^2+ (k\sigma_0^2/2)^4)^{1/2}}{k\sigma_0^2/2}=\frac{\sigma_z}{\lambda},
\end{align}
where $\tilde{\sigma}_z$ is the scaled, $\tilde{z}-$parameterized standard deviation of the model, $\sigma_0$ is the un-scaled standard deviation at $z=\tilde{z}=0$, and $\sigma_z$ is the un-scaled, $z$-dependent standard deviation.  
\par
\begin{figure}[tbh]
 \centerline{
  \begin{tabular}{c}
  \includegraphics[width=9cm,angle=0]{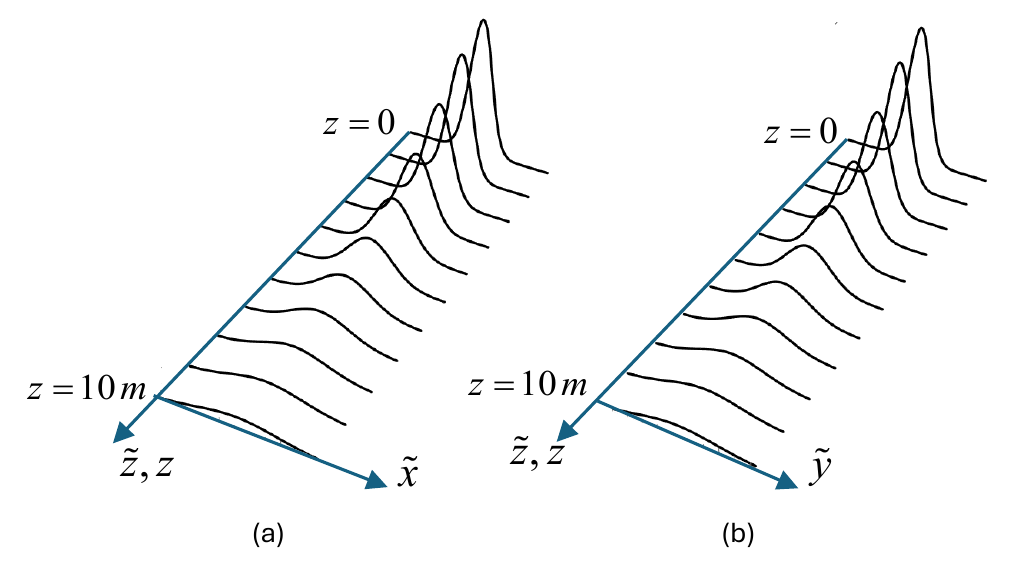}   
  \end{tabular}
  }
   \caption{Propagation of the intensity of a circularly-symmetric Gaussian beam at $\lambda=1.55\,\mu m$ for $\tilde{\sigma}_{0}=1.25\,mm$. (a) Intensity profiles in the $\tilde{x}\tilde{z}-$plane at $\tilde{y}=0$. (b) Intensity profiles in the $\tilde{y}\tilde{z}-$plane at $\tilde{x}=0$. }
   \label{fig:CGSlicesSummary}   
\end{figure}

 Figure \ref{fig:CGSlicesSummary} shows a schematic representation of the propagation of the beam intensity at a few select $z-$distances from the source at $z=\tilde{z}=0$. Shown are both "$x-$slices" and "$y-$slices".

\par As required, $\rho_{00}$ with scaled coordinates and scaled parameter is normalized over the transverse plane
\begin{align}\label{eq:Rho00Normalization}
\int_{-\infty}^{\infty}\int_{-\infty}^{\infty}
\rho_{00}\left(\tilde{x}, \tilde{y}|\,\tilde{\sigma}_{z}
\right)d\tilde{x}d\tilde{y}=1,
\end{align}
independent of $\tilde{z}$.

For the probability density (\ref{eq:Rho00}), 
\begin{align}
\ln\rho_{00}=\ln\left(\dfrac{2}{\pi}\right)-2\ln\left({\tilde{\sigma}_z}\right)-\dfrac{2(\tilde{x}^2+\tilde{y}^2)}{\tilde{\sigma}_z^2}.
\end{align}
Since (\ref{eq:Rho00})  is a one-parameter model the Hessian comprises just a single term,
\begin{align}\label{eq:HessianRho00}
{\bf{H}}_{00}(\tilde{x},\tilde{y}|\,\tilde{\sigma}_z)&=
\dfrac{\partial^2}{\partial{\tilde{\sigma_z}^2}}\ln \rho_{00}(\tilde{x},\tilde{y}\,|\,\tilde{\sigma}_z)
=\dfrac{2}{\tilde{\sigma}_z^2}-\dfrac{12(\tilde{x}^2+\tilde{y}^2)}{\tilde{\sigma}_z^4}
.
\end{align}
The optical pressure tensor is
\begin{align}
{\bf{P}}_{00}(\tilde{x},\tilde{y})&=\dfrac{-\rho_{00}(\tilde{x},\tilde{y})}{4(2\pi)^2}\begin{bmatrix}
\partial^2/\partial \tilde{x}^2 & \partial^2/\partial \tilde{x}\partial \tilde{y} \\
\partial^2/\partial \tilde{y}\partial \tilde{x} & \partial^2/\partial \tilde{y}^2
\end{bmatrix}\ln\left(\rho_{00}(\tilde{x},\tilde{y})\right),
\nonumber\\[10pt]
&=\dfrac{\tilde{\rho}_{00}(\tilde{x},\tilde{y},\tilde{z}|\,\tilde{\sigma}_z)}{2\pi^2\tilde{\sigma}_z^2}\begin{bmatrix} 1 & 0\\ 0 & 1
\end{bmatrix},
\end{align}
and the corresponding transverse acceleration vector is
\begin{align}  \label{eq:GaussAccel}
\tilde{\vec{a}}_{00}(\tilde{x},\tilde{y})=\dfrac{-1}{\rho_{00}(\tilde{x},\tilde{y})}\nabla_T\cdot{\bf{P}}=\dfrac{1}{\pi^2\tilde{\sigma}_z^4}\begin{bmatrix}\tilde{x} \\\tilde{y}
\end{bmatrix}.
\end{align}
The tensor ${\bf{P}}_{00}$ and the vector $\vec{a}_{00}$ are both represented in the standard Cartesian basis.

\begin{figure}[t]
 \centerline{
  \begin{tabular}{c}
  \includegraphics[width=8cm,angle=0]
  {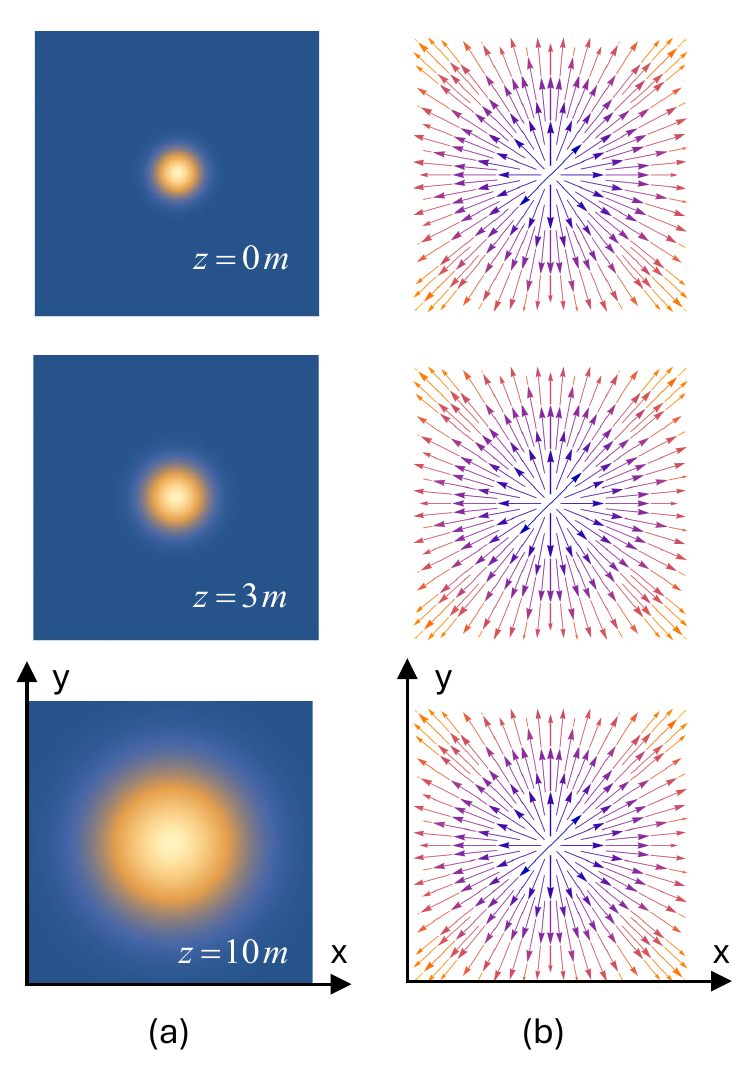}   
  \end{tabular}
  }
   \caption{(a) Density plots of the intensity pdf  for the circularly-symmetric Gaussian beam (\ref{eq:BVDensity}) at the indicated $z-$ positions for model parameters $\lambda=1.55\mu m, \sigma_{0}=1.25\,mm$. Here each intensity plot has been normalized to the value at the origin. (b) Corresponding acceleration vector field depicted as streamlines. Each plot has dimensions $-4000\,m/m\leq\tilde{x},\tilde{y}\leq4000\,m/m$}
   \label{fig:CGDensityAccel}   
\end{figure}

Figure~\ref{fig:CGDensityAccel} shows plots of the acceleration streamlines for the corresponding intensity pdf of the circularly-symmetric Gaussian beam at three different downrange positions $z=0, 3, 10\,m$. As predicted in section \ref{sec:minFIM}, the movement of intensity is clearly away from the maximum and occurs at a rate that increases with transverse distance from the origin.
  As $\tilde{z}\rightarrow\infty$, $\tilde{\sigma}_z^2\rightarrow\infty$, and the intensity distribution flattens -- consistent with the well-known diffraction properties of a Gaussian beam. Again, keep in mind that that the distribution cannot become uniformly flat over the transverse plane since the distribution must remain integrable (Eq.~(\ref{eq:Rho00Normalization})).

The associated FIM and ${\bf{F}}$ are given by
\begin{subequations}
\begin{align}
{\mathcal{F}}_{00}(\tilde{\sigma_z})&=\frac{4}{\tilde{\sigma}_z^2},
\label{eq:CGF}\\
{\bf F}_{00}(\tilde{\sigma}_z)&=\det\Big({\mathcal{F}}(\tilde{\sigma_z})\Big)=\frac{4}{\tilde{\sigma}_z^2}.
\end{align}
\end{subequations}
In this case, where there is a single model parameter $\tilde{\sigma}_z$, $\mathcal{F}$ is a $1\times1$ matrix and, hence, $\mathcal{F}$ and ${\bf{F}}$ have the same value.  As $\tilde{z}\rightarrow\infty$, $\tilde{\sigma}_{z}^2\rightarrow \tilde{z}^2/\pi^2\tilde{\sigma}^2_{0}$ and so ${\bf F}_{00}\rightarrow {\bf 0}$. Thus the physics of diffraction are indeed associated with minimization of the Fisher information for this Gaussian intensity model.  

Finally, the rate of change of ${\bf{F}}_{00}$ is given by 
\begin{align}
    \partial_{\tilde{z}}\,{\bf F}_{00}&=\frac{ -8\,\tilde{z}}{\pi^2\sigma_0^2\,\tilde{\sigma}_z^2}.
\end{align}
The rate of change of FI is therefore always negative and, hence, the decrease in FI is monotonic with propagation distance.  Plots of ${\bf{F}}_{00}$, and $\partial_{\tilde{z}}{\bf{F}}_{00}$ are presented and discussed in Section~\ref{sec:Disc}.

\subsection{Uniformly-polarized Hermite-Gaussian Beam}
Hermite-Gaussian (HG) beams comprise a complete set of solutions to the PWE.  The intensity of the lowest order HG beam is simply the circularly-symmetric Gaussian beam of Eq.~(\ref{eq:Rho00}). Higher-order beam modes are identified by two integer indices, denoted $n$ and $m$. The intensity of the $\{n,m\}$ Hermite-Gaussian mode at the source is
\begin{align}\label{eq:HGmode}
\rho_{nm}(\tilde{x},\tilde{y},0|\,\tilde{\sigma}_0,n,m)={\mathcal{H}_{nm}(\tilde{x},\tilde{y},\tilde{z}|\,\tilde{\sigma}_0)}\rho_{00}(\tilde{x},\tilde{y},0|\tilde{\sigma}_0) 
\end{align}
where
\begin{align}\label{eq:DefineHermite}
{\mathcal{H}_{nm}}(\tilde{x},\tilde{y},0|\,\tilde{\sigma}_0)=\dfrac{1}{2^nn!2^mm!}H^2_n\left(\dfrac{\sqrt{2}\,\tilde{x}}{\tilde{\sigma}_0}\right)H^2_m\left(\dfrac{\sqrt{2}\,\tilde{y}}{\tilde{\sigma_0}}\right)
\end{align}
and where $H_n(\cdot)$ denotes the Hermite polynomial of order $n$.

In this example we consider the HG beam $\rho_{10}$. Since $H_0(u)=1$ and $H_1(u)=2u$, then
\begin{align}\label{eq:HG10}
\rho_{10}(\tilde{x},\tilde{y},0|\,\tilde{\sigma}_0)=\dfrac{8\tilde{x}^2}{\pi\tilde{\sigma}_0^4}\exp{\left(-2(\tilde{x}^2+\tilde{y}^2)/\tilde{\sigma}_0^2\right)}, 
\end{align}  
and, as required, $\rho_{10}(\tilde{x},\tilde{y},0|\,\tilde{\sigma}_0)$ is normalized over the transverse plane, 
\begin{align}
\int_{-\infty}^{\infty}\int_{-\infty}^{\infty}\rho_{10}(\tilde{x},\tilde{y},\tilde{z}|\,\tilde{\sigma}_z)d\tilde{x}d\tilde{y}=1.
\end{align}

Following the same procedure as in Section 5.1 for the circularly-symmetric Gaussian beam, the z-dependent probability density for the $\rho_{10}$ beam is obtained by first assuming a probability density at the source plane $(z=\tilde{z}=0)$ given by (\ref{eq:HG10}) with $(\tilde{x},\tilde{y})=(\tilde{x}_1,\tilde{y}_1)$ and for fixed standard deviation $\tilde{\sigma}_0$ and then assuming further that the corresponding scalar electric field $A_{10}(\tilde{x_1},\tilde{y_1},0|\,\tilde{\sigma}_0)$ is given by
\begin{align}
A_{10}(\tilde{x}_1,\tilde{y}_1,0|\,\tilde{\sigma}_0)=\Bigg(\dfrac{8}{\pi\tilde{\sigma}_0^4}\Bigg)^{1/2}\,\tilde{x}_1\,\exp{\Big(-(\tilde{x}_1^2+\tilde{y}_1^2)/\tilde{\sigma}_0^2\Big)}.
\end{align}
The field $A_{10}(\tilde{x_1},\tilde{y_1},0|\,\tilde{\sigma}_0)$ is then propagated along the positive $\tilde{z}-$axis using the Huygens-Fresnel integral (\ref{eq:HFIntegral}) to obtain 
\begin{align}
A_{10}(\tilde{x},\tilde{y},\tilde{z}|\,\tilde{\sigma}_0)=\dfrac{2\sqrt{2}\pi^{3/2}\tilde{\sigma}_0^2}{\left(\pi\tilde{\sigma}_0^2-i\tilde{z}\right)^2}\,\tilde{x}\,\exp{\Bigg(\dfrac{-\pi(\tilde{x}^2+\tilde{y}^2)}{\pi\tilde{\sigma}_0^2-i\tilde{z}}\Bigg)}.
\end{align}
Finally, the probability density corresponding to the $z-$dependent intensity distribution is
$
\rho_{10}(\tilde{x},\tilde{y},\tilde{z}|\,\tilde{\sigma}_0)=|A_{10}(\tilde{x},\tilde{y},\tilde{z}|\,\tilde{\sigma}_0)|^2
$
which yields 
\begin{align}\label{eq:HGDownrangeIntensity}
\rho_{10}(\tilde{x},\tilde{y},\tilde{z}|\,\tilde{\sigma}_0)=\dfrac{8\pi^3\tilde{\sigma}_0^4}{\left(\pi^2\tilde{\sigma}_0^4+\tilde{z}^2\right)^2}\tilde{x}^2\exp{\Bigg(\dfrac{-2\pi^2\tilde{\sigma}_0^2(\tilde{x}^2+\tilde{y}^2)}{\pi^2\tilde{\sigma}_0^4+\tilde{z}^2}\Bigg)}.
\end{align}

\par
\begin{figure}[tbh]
 \centerline{
  \begin{tabular}{c}
  \includegraphics[width=9cm,angle=0]{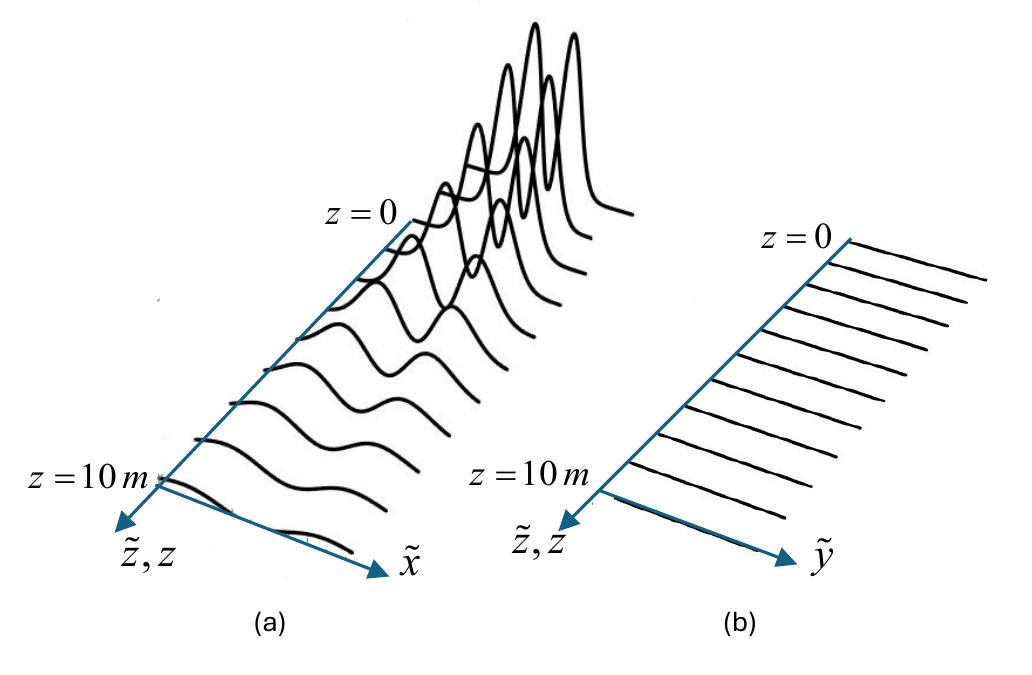}   
  \end{tabular}
  }
   \caption{Propagation of a Hermite-Gaussian beam at $\lambda=1.55\,\mu m$ for $\sigma_{0}=1.25\,mm$. (a) Intensity profiles in the $\tilde{x}\tilde{z}-$plane at $\tilde{y}=0$. (b) Intensity profiles in the $\tilde{y}\tilde{z}-$plane at $\tilde{x}=0$. Note that in this case the initial intensity vanishes along the $\tilde{y}-$axis and the intensity remains zero along this axis as the beam propagates. }
   \label{fig:HGSlicesSummary}   
\end{figure}

 Figure \ref{fig:HGSlicesSummary} shows a schematic representation of the propagation of the beam at a few select $z-$distances from the source at $z=\tilde{z}=0$. Shown are both "$x-$slices" and "$y-$slices".

Upon propagation the beam spreads in the transverse plane and the intensity flattens but never becomes completely flat in the sense that the normalization 
\begin{align}
\int_{-\infty}^{\infty}\int_{-\infty}^{\infty}\rho_{10}(\tilde{x},\tilde{y},\tilde{z}|\,\tilde{\sigma}_0)d\tilde{x}d\tilde{y}=1
\end{align}
is satisfied for all $\tilde{z}$. 

Using (\ref{eq:sigz}), the downrange intensity pdf (\ref{eq:HGDownrangeIntensity}) can be written instead with $\tilde{\sigma}_z$ as the model parameter,
\begin{align}\label{eq:HGDownrangeIntensitySigz}
\rho_{10}(\tilde{x},\tilde{y},\tilde{z}|\,\tilde{\sigma}_z)=\dfrac{8\tilde{x}^2}{\pi\tilde{\sigma}^4_z}\exp{\Bigg(\dfrac{-2(\tilde{x}^2+\tilde{y}^2)}{\tilde{\sigma}_z^2}\Bigg)},
\end{align}
and then

\begin{align}
\ln\big({\rho_{10}(\tilde{x},\tilde{y},\tilde{z}|\tilde{\sigma}_0)}\big)=\ln{\Bigg(\dfrac{8}{\pi}\Bigg)}+2\ln{\tilde{x}}-4\ln\tilde{\sigma}_z-2(\tilde{x}^2+\tilde{y}^2)/\tilde{\sigma}_z^2.
\end{align}
For the HG$_{10}$ beam with one model parameter $\tilde{\sigma}_z$ we find, using the expressions in Table 1,
\begin{subequations}\label{eq:HGSummary}
\begin{align}
{\bf H}_{10}(\tilde{x},\tilde{y}\,\tilde{z}|\,\tilde{\sigma}_z)&=
\dfrac{4}{\tilde{\sigma}_z^2}-\dfrac{12(\tilde{x}^2+\tilde{y}^2)}{\tilde{\sigma}_z^4},\label{eq:HessianHG01}
\\[4pt]
{\bf{P}}_{10}(\tilde{x},\tilde{y})&=\dfrac{\exp{\left(-2(\tilde{x}^2+\tilde{y}^2)/\sigma_z^2\right)}}{\pi^3}\begin{bmatrix} \dfrac{1}{\tilde{\sigma}_z^4}+\dfrac{2\tilde{x}^2}{\tilde{\sigma}_z^6} & 0\\ 0 & \dfrac{2\tilde{x}^2}{\tilde{\sigma}_z^6}
\end{bmatrix},
\\[4pt]
\vec{a}_{10}(\tilde{x},\tilde{y})&=\dfrac{1}{\pi^2\tilde{\sigma}_z^4}\begin{bmatrix}
x \\ y
\end{bmatrix},
\\[4pt]
{\mathcal{F}}_{10}(\tilde{\sigma}_z)&={\bf{F}}_{10}(\tilde{\sigma}_z)=
\dfrac{8}{\tilde{\sigma}_z^2},\label{eq:HGF}\\[5pt]
\partial_{\tilde{z}}{\bf{ F}}_{10}(\tilde{\sigma}_z)&= \dfrac{-16}{\pi^2\sigma_0^2}\dfrac{\tilde{z}}{\tilde{\sigma}_z^2}
\label{eq:decay}
\end{align}
\end{subequations}

\begin{figure}[ht]
 \centerline{
  \begin{tabular}{c}
  \includegraphics[width=8cm,angle=0]
  {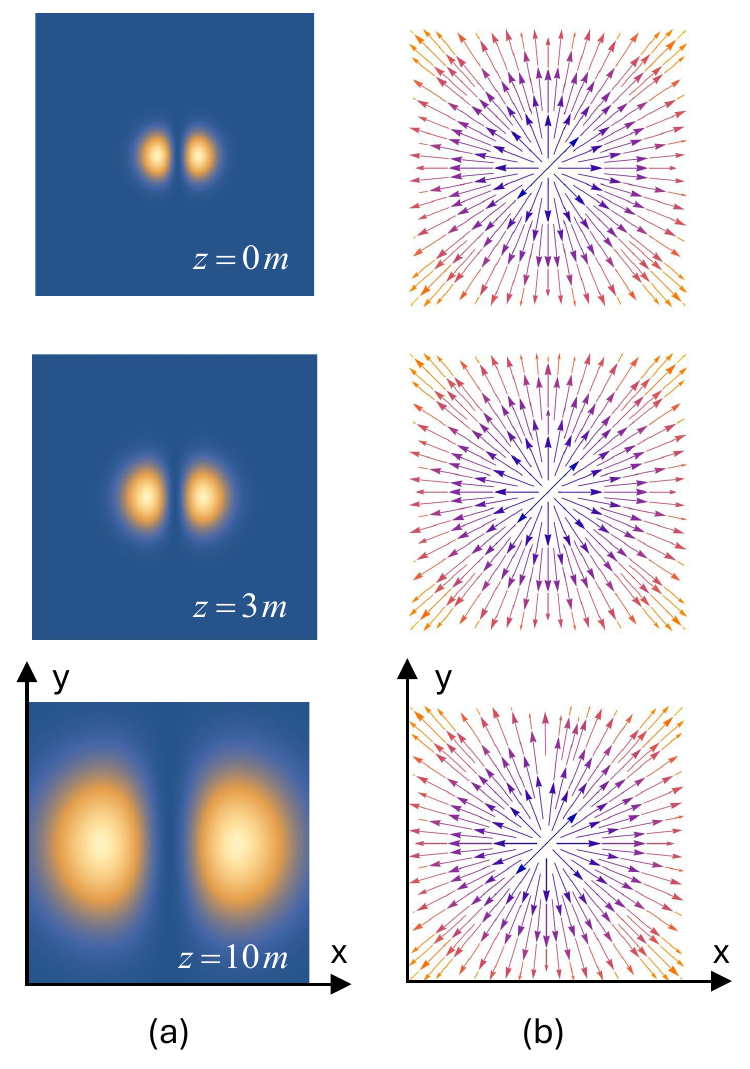}   
  \end{tabular}
  }
   \caption{(a) Density plots of the intensity pdf  for the 10-Hermite Gaussian beam (\ref{eq:BVDensity}) at the indicated $z-$ positions for model parameters $\lambda=1.55\mu m,  \sigma_{0}=1.25\,mm$. Here each intensity plot has been normalized to the value at the origin. (b) Corresponding acceleration vector field depicted as streamlines. Each plot has dimensions $-4000\,m/m\leq\tilde{x},\tilde{y}\leq4000\,m/m$}
   \label{fig:HGDensityAccel}   
\end{figure}

Figure~\ref{fig:HGDensityAccel} shows plots of the acceleration streamlines for the corresponding the intensity pdf of the propagating Hermite-Gaussian beam at three different downrange positions $z=0, 3, 10\,m$. Again we see from Eq. (\ref{eq:decay}) that the decrease in FI is monotonic on propagation. 

Plots of both ${\bf{F}}_{10}$, and $\partial_z{\bf{F}}_{10}$ are presented and discussed below in Section~\ref{sec:Disc}.

\subsection{Uniformly-polarized Bivariate Gaussian Beam}

As a final example consider a uniformly-polarized, bivariate Gaussian distribution centered on the origin where, in general, the probability density  $\rho_{bv}$ can be written
\begin{align}\label{eq:BVCovariance}
\rho_{bv}(\tilde{x},\tilde{y},0|\tilde{\sigma}_{0x},\tilde{\sigma}_{0y},r)= \dfrac{2}{\pi\sqrt{\det(\widetilde{\mathcal{S}}_0})}\exp{\big(-2{\tilde{\bf{x}}}^T\widetilde{\mathcal{S}}_0^{-1}{\tilde{\bf{x}}}\big)}
\end{align}
where the initial covariance matrix $\widetilde{\mathcal{S}}_0$, given by
\begin{align}
\widetilde{\mathcal{S}}_0=\begin{bmatrix} \tilde{\sigma}_{0x}^2 & r\tilde{\sigma}_{0x}\tilde{\sigma}_{0y} \\
r\tilde{\sigma}_{0x}\tilde{\sigma}_{0y} & \tilde{\sigma}_{0y}^2
\end{bmatrix},
\end{align}
is characterized by three model parameters: standard deviations $\tilde{\sigma}_{0x},\,\tilde{\sigma}_{0y}$ that describes widths of the distribution in the $\tilde{x}-$ and $\tilde{y}-$directions, respectively, and correlation coefficient $-1< r < +1$ which determines the orientation of the distribution in the $\tilde{x}\tilde{y}-$plane. The position vector in the transverse plane is ${\tilde{\bf{x}}}=[\tilde{x}\,\,\tilde{y}]^T$ and $\widetilde{\mathcal{S}}_0^{-1}$ is the matrix inverse of $\widetilde{\mathcal{S}}_0$. Note that the cases $r=\pm 1$ correspond to a univariate Gaussian distribution along the lines $\tilde{y}=-\tilde{x}$ and $\tilde{y}=\tilde{x}$, respectively, and hence these two cases are excluded from the definition of the bivariate distribution. Written explicitly in terms of the three model parameters,
\begin{align}\label{eq:BVStandard}
\rho_{bv}(\tilde{x},\tilde{y},0|\tilde{\sigma}_{0x},\tilde{\sigma}_{0y},r)= \dfrac{2}{\pi\tilde{\sigma}_{0x}\tilde{\sigma}_{0y}\sqrt{1-r^2}}\exp\Bigg[{\dfrac{-2}{(1-r^2)}\Bigg(\dfrac{\tilde{x}^2}{\tilde{\sigma}_{0x}^2}+\dfrac{\tilde{y}^2}{\tilde{\sigma}_{0y}^2}-\dfrac{2r\tilde{x}\tilde{y}}{\tilde{\sigma}_{0x}\tilde{\sigma}_{0y}}\Bigg)}\Bigg].
\end{align}

In this section we consider a beam in which the \emph{initial} intensity distribution has the form (\ref{eq:BVStandard}). As in the previous two examples, the initial electric field is taken to be the square root of the initial intensity,
\begin{align}
A_{bv}(\tilde{x},\tilde{y},0|\tilde{\sigma}_{0x},\tilde{\sigma}_{0y},r)=\Bigg(\dfrac{2}{\pi\tilde{\sigma}_{0x}\tilde{\sigma}_{0y}\sqrt{1-r^2}}\Bigg)^{1/2}\exp\Bigg[{\dfrac{-1}{(1-r^2)}\Bigg(\dfrac{x^2}{\tilde{\sigma}_{0x}^2}+\dfrac{y^2}{\tilde{\sigma}_{0y}^2}-\dfrac{2rxy}{\tilde{\sigma}_{0x}\tilde{\sigma}_{0y}}\Bigg)}\Bigg].
\end{align}
Application of the HF integral yields the propagating field
\begin{align}
A_{bv}(\tilde{x},\tilde{y},\tilde{z}|\tilde{\sigma}_{0x},\tilde{\sigma}_{0y},r)=\dfrac{i}{z}\sqrt{\dfrac{2}{B_2(\tilde{x},\tilde{y},\tilde{z},\tilde{\sigma}_{0x},\tilde{\sigma}_{0y},r)\sqrt{1-r^2}}}\exp{\Big[-B_1(\tilde{x},\tilde{y},\tilde{z},\tilde{\sigma}_{0x},\tilde{\sigma}_{0y},r)\Big]}
\end{align}
where
\begin{subequations}
\begin{align}
B_1(\tilde{x},\tilde{y},\tilde{z},\tilde{\sigma}_{0x},\tilde{\sigma}_{0y},r)&=\dfrac
{\pi\Big[\pi\Big(\tilde{\sigma}_{0y}^2\tilde{x}^2-2r\tilde{\sigma}_{0x}\tilde{\sigma}_{0y}\tilde{x}\tilde{y}+\tilde{\sigma}_{0x}^2\tilde{y}^2\Big)-i(\tilde{x}^2+\tilde{y}^2)\tilde{z}\Big]}
{\pi^2(1-r^2)\tilde{\sigma}_{0x}^2\tilde{\sigma}_{0y}^2+i\pi(\tilde{\sigma}_{0x}^2+\tilde{\sigma}_{0y}^2)\tilde{z}+\tilde{z}^2},
\\
B_2(\tilde{x},\tilde{y},\tilde{z},\tilde{\sigma}_{0x},\tilde{\sigma}_{0y},r)&=\Bigg(\dfrac{1}{\pi}+\dfrac{r^2\tilde{\sigma}_{0y}^2}{\pi(1-r^2)\tilde{\sigma}_{0y}^2+\dfrac{i\tilde{\sigma}_{0x}^2}{z}}\Bigg)\Bigg(\dfrac{1}{\tilde{\sigma}_{0x}\tilde{\sigma}_{0y}(1-r^2)}+\dfrac{i\pi\tilde{\sigma}_{0x}}{\tilde{z}\tilde{\sigma}_{0x}}\Bigg),
\end{align}
\end{subequations}
and the downrange intensity pdf is
\begin{align}
\rho_{bv}(\tilde{x},\tilde{y},\tilde{z}|\,\tilde{\sigma}_{0x},\tilde{\sigma}_{0y},r)&=|A_{bv}(\tilde{x},\tilde{y},\tilde{z}|\tilde{\sigma}_{0x},\tilde{\sigma}_{0y},r)|^2.
\end{align}

We will see that, upon propagation, the pdf of the beam's intensity no longer has the simple form of (\ref{eq:BVStandard}) because, due to diffraction, the two variances change with scaled propagation distance $\tilde{z}$ at different rates, in general. As a result, for $\tilde{z}>0$, the variance of the cross term is no longer simply the product of the other two variances but instead must be treated as a unique, third parameter. 
The propagated intensity pdf can be written 

 \begin{align}  \label{eq:BVDensity}
 \rho_{bv}(\tilde{x},\tilde{y}\,|\,\tilde{\sigma}_{zxx},\tilde{\sigma}_{zyy},\tilde{\sigma}_{zxy})=
 \left(\frac{2}{\pi\sqrt{\det(\widetilde{\mathcal{S}}_z)}}\right)\exp{\Big(-2(\tilde{\vec{x}})^T\widetilde{\mathcal{S}}^{-1}_z\tilde{\vec{x}}\Big)},
\end{align}
where the $z-$dependent covariance matrix is given by
\begin{align}
\widetilde{\mathcal{S}_z}&=\left[\begin{array}{cc}
\tilde{\sigma}_{zxx}^2 & -r\tilde{\sigma}_{zxy}^2\\[8pt]
-r\tilde{\sigma}_{zxy}^2 & \tilde{\sigma}_{zyy}^2
\end{array}
\right],
\end{align}
and where\footnote{It is easy to show that $\widetilde{\mathcal{S}_z}$ reduces to $\widetilde{\mathcal{S}}_0$ at $\tilde{z}=0.$}
\begin{subequations}\label{eq:BVvariances}
\begin{align}
\tilde{\sigma}^2_{zxx}&= \left(\frac{\tilde{z}^2}{(2\pi)^2(1-r^2)\tilde{\sigma}_{0y}^2}+\tilde{\sigma}_{0x}^2\right)
= \dfrac{1}{\lambda^2}\left(\frac{4z^2}{k^2(1-r^2)\sigma_{0x}^2}+\sigma_{0x}^2\right),
\\[8pt]
\tilde{\sigma}^2_{zyy}&= \left(\frac{\tilde{z}^2}{(2\pi)^2(1-r^2)\tilde{\sigma}_{0y}^2}+\tilde{\sigma}_{0y}^2\right)
= \dfrac{1}{\lambda^2}\left(\frac{4z^2}{k^2(1-r^2)\sigma_{0y}^2}+\sigma_{0y}^2\right),
\\[8pt]
\tilde{\sigma}^2_{zxy}&=\left(\frac{\tilde{z}^2}{(2\pi)^2(1-r^2)\tilde{\sigma}_{0x}\tilde{\sigma}_{0y}}-\tilde{\sigma}_{0x}\tilde{\sigma}_{0y}\right)=\dfrac{1}{\lambda^2}\left(\frac{4z^2}{k^2(1-r^2)\sigma_{0x}\sigma_{0y}}+\sigma_{x0}\sigma_{0y}\right).
\end{align}
\end{subequations}
 Recall that $\tilde{\sigma}^2_{0x}$ and $\tilde{\sigma}^2_{0y}$ are the initial $(\tilde{z}=0)$ variances and that a tilde over a quantity indicates that the quantity has been scaled by wavelength as discussed above in the context of (\ref{eq:ScaledQuantities}). For completeness, in Eqs. (\ref{eq:BVvariances}) we show both the "tilded" and the "un-tilded" version of each variance. Importantly, note that for $\tilde{z}>0$,   $\tilde{\sigma}^2_{zxy}\neq \tilde{\sigma}_{zxx}\tilde{\sigma}_{zyy}$. 
\emph{Hence, propagation of an optical beam with initial bivariate-Gaussian intensity pdf must be described by the more general form (\ref{eq:BVDensity}) rather than the standard form (\ref{eq:BVStandard}).}
\par
\begin{figure}[tbh]
 \centerline{
  \begin{tabular}{c}
  \includegraphics[width=9cm,angle=0]
  {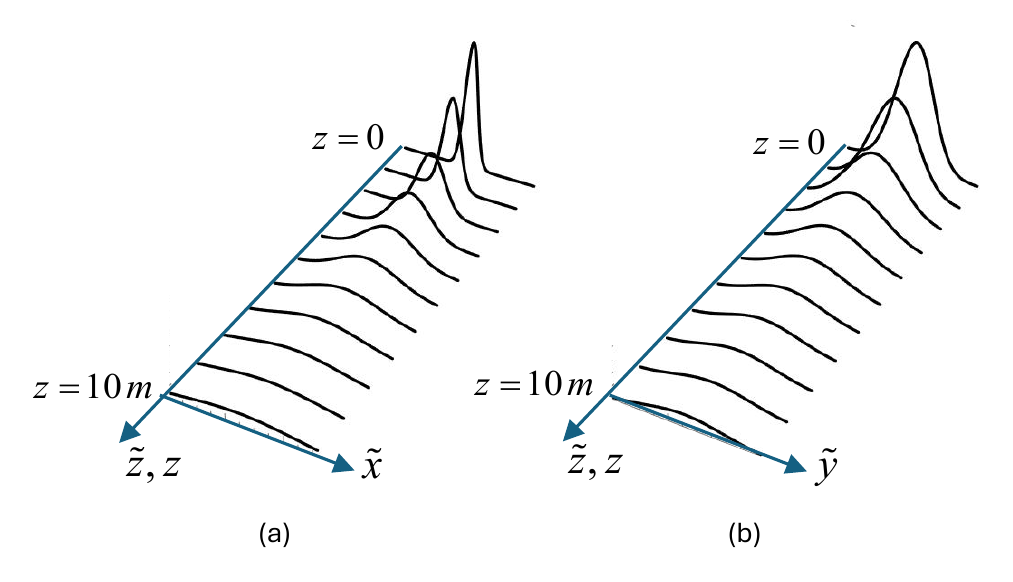}   
  \end{tabular}
  }
   \caption{Propagation of a bivariate Gaussian beam at $\lambda=1.55\,\mu m$ for model parameters $\sigma_{0x}=1.25\,mm,\, \sigma_{0y}=3.75\,mm,\,r=0.8$. (a) Intensity profiles in the $\tilde{x}\tilde{z}-$plane at $\tilde{y}=0$. (b) Intensity profiles in the $\tilde{y}\tilde{z}-$plane at $\tilde{x}=0$. Note that the rate of diffractive spreading is larger in the transverse direction with the smaller initial variance, in this case, the $\tilde{x}-$direction. }
   \label{fig:BVSlicesSummary}   
\end{figure}

 Figure \ref{fig:BVSlicesSummary} shows a schematic representation of propagation of the bivariate beam at a few select $z-$distances from the source. Shown are both "$x-$slices" and "$y-$slices".

It is important to note again that the entire procedure described here was possible because the beam was assumed to be uniformly polarized. For non-uniformly polarized light this approach is not applicable as the vector components of the electric field cannot be treated as independent (see, for example, Nichols {\it et al.} \cite{Nichols:24}).

As before, we calculate
\begin{align}
\ln{\rho_{bv}}=\ln{\left(\dfrac{2}{\pi}\right)}-(1/2)\ln{\left(\det{(\widetilde{\mathcal{S}})}\right)}-2\left(\tilde{\vec{x}}\right)^T\widetilde{S}^{-1}\tilde{\vec{x}}.
\end{align}
from which, for the bivariate Gaussian beam characterized by the three parameters $[\begin{matrix}\theta_1, & \theta_2, & \theta_3\end{matrix}]\equiv [\begin{matrix}\tilde{\sigma}_{zxx}, & \tilde{\sigma}_{zyy}, & \tilde{\sigma}_{zxy}\end{matrix}]$. Continuing we calculate
\begin{subequations}
\begin{align}
H_{bv}(i,j)(\tilde{x},\tilde{y}\,|\,\tilde{\sigma}_{zxx},\,\tilde{\sigma}_{zyy},\,\tilde{\sigma}_{zxy})&=\dfrac{\partial^2}{\partial\theta_i\partial\theta_j}\ln \rho_{bv}(\tilde{x},\tilde{y}\,|\,\tilde{\sigma}_{zxx},\,\tilde{\sigma}_{zyy},\,\tilde{\sigma}_{zxy}),\quad i,j\in\{1,2,3\},\label{eq:BVHessian}\\
{\bf{P}}_{bv}(\tilde{x},\tilde{y})&=\dfrac{-\rho_{bv}(\tilde{x},\tilde{y})}{4(2\pi)^2}\left[\begin{array}{cc} \dfrac{\partial^2 }{\partial \tilde{x}^2} & \dfrac{\partial^2 }{\partial \tilde{x}\partial \tilde{y}}\\
\dfrac{\partial^2 }{\partial \tilde{y}\partial \tilde{x}} & \dfrac{\partial^2 }{\partial \tilde{y}^2}
\end{array}
\right]\ln\left(\rho_{bv}(\tilde{x},\tilde{y}\right))\\
\vec{a}_{bv}(\tilde{x},\tilde{y})&=\dfrac{-1}{\rho_{bv}}\tilde{\nabla}_T\cdot{\bf{P}}_{bv},\label{eq:BVAccel}
\end{align}
\end{subequations}

\begin{figure}[t]
 \centerline{
  \begin{tabular}{c}
  \includegraphics[width=8cm,angle=0]
  {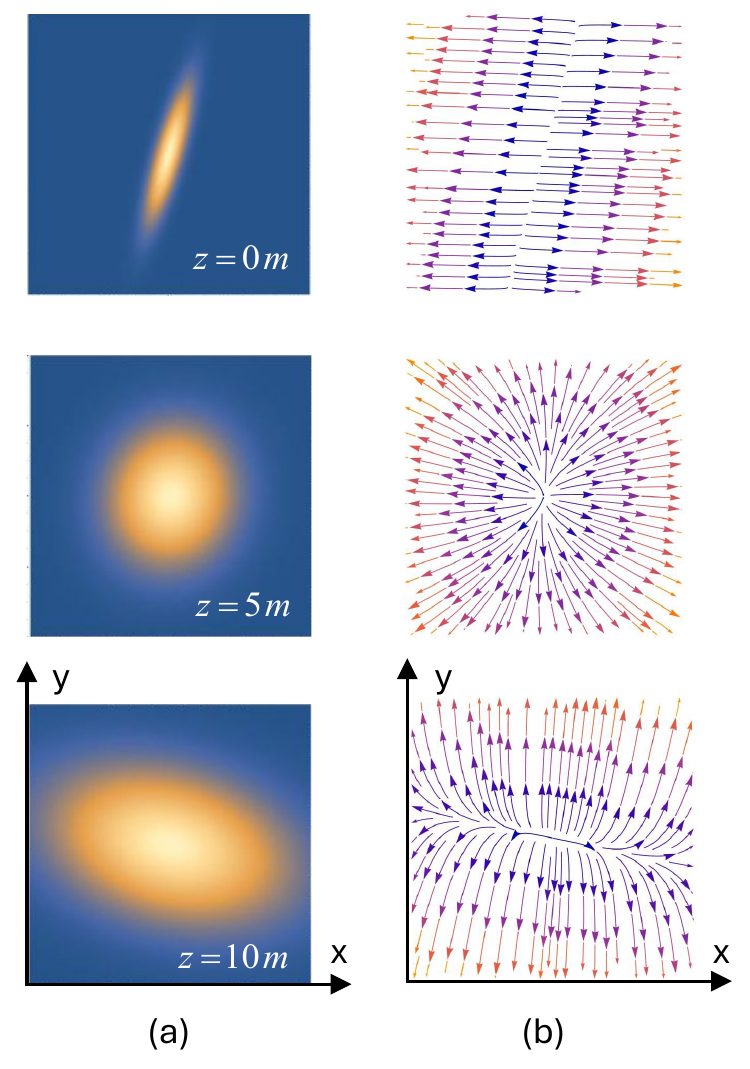}   
  \end{tabular}
  }
   \caption{(a) Density plots of the intensity pdf  for the bivariate Gaussian beam (\ref{eq:BVDensity}) at the indicated $z-$ positions for model parameters $\lambda=1.55\mu m, \sigma_{0x}=1.25\,mm,\,\sigma_{0y}=3.75\,mm,r=0.8$. (b) Corresponding acceleration vector field depicted as streamlines. Each plot has dimensions $-4000\,m/m\leq\tilde{x},\tilde{y}\leq4000\,m/m$}
   \label{fig:BVDensityAccel}   
\end{figure}

Explicit expressions for $H_{bv}(i,j),{\bf{P}}_{bv}$ and $\vec{a}_{bv}$ are long, cumbersome, and not particularly informative. They can be obtained using a variety of available methods.\footnote{We used Mathematica for all calculations in this paper.}  However, it is quite useful to examine the movement of optical intensity graphically by plotting the acceleration streamlines along with the associated intensity profiles as a function of propagation distance. Figure~\ref{fig:BVDensityAccel} shows plots of the acceleration streamlines corresponding to the intensity pdf of the propagating bivariate Gaussian beam at three different downrange positions $z=0, 5, 10\,m$. 

The Fisher information matrix for this beam is the symmetric $3\times 3$ matrix 

 \begin{align}
 &{\mathcal{F}}_{bv}(\tilde{\sigma}_{zxx},\tilde{\sigma}_{zyyx},\tilde{\sigma}_{zxy})=\\
 &\qquad \dfrac{1}{D^2}\begin{bmatrix}
2\tilde{\sigma}^2_{zxx}\tilde{\sigma}^4_{zyy} & 
2r^2\tilde{\sigma}_{zxx}\tilde{\sigma}_{zyy}\tilde{\sigma}^4_{zxy}&
-4r^2\tilde{\sigma}_{zxx}\tilde{\sigma}^2_{zyy}\tilde{\sigma}^3_{zxy}\\[10pt]
2r^2\tilde{\sigma}_{zxx}\tilde{\sigma}_{zyy}\tilde{\sigma}^4_{zxy} &
2\tilde{\sigma}^4_{zxx}\tilde{\sigma}^2_{zyy}&
-4r^2\tilde{\sigma}_{zxx}\tilde{\sigma}^2_{zyy}\tilde{\sigma}^3_{zxy}\\[10pt]
-4r^2\tilde{\sigma}_{zxx}\tilde{\sigma}^2_{zyy}\tilde{\sigma}^3_{zxy} & 
-4r^2\tilde{\sigma}_{zxx}\tilde{\sigma}^2_{zyy}\tilde{\sigma}^3_{zxy}& 
4r^2\tilde{\sigma}^2_{zxy}\left(\tilde{\sigma}^2_{zxx}\tilde{\sigma}^2_{zyy}+r^2\tilde{\sigma}^4_{zxy}\right)
 \end{bmatrix}.
 \end{align}
where
\begin{align}
D=r^2\tilde{\sigma}_{zxy}^4-\tilde{\sigma}_{zxx}^2\tilde{\sigma}_{zyy}^2.
\end{align}

 The Fisher information
 \begin{align}
 {\bf{F}}_{bv}(\tilde{\sigma}_{zxx},\tilde{\sigma}_{zyyx},\tilde{\sigma}_{zxy})&=\dfrac{-16 r^2\tilde{\sigma}^2_{zxx}\tilde{\sigma}^2_{zyy}\tilde{\sigma}^2_{zxy}}{D^6}\Bigg(
r^6\tilde{\sigma}^{12}_{zxy}+r^4\tilde{\sigma}_{zxx}\tilde{\sigma}^8_{zxy}\left(\tilde{\sigma}_{zxx}-4\tilde{\sigma}_{zyy}\right)\tilde{\sigma}^2_{zyy}\nonumber\\[10pt]
&\qquad\qquad\qquad\qquad\qquad\quad -\tilde{\sigma}^6_{zxx}\tilde{\sigma}^6_{yy}\nonumber\\[10pt]
&\qquad\qquad\qquad\qquad\qquad\quad\quad
+r^2\tilde{\sigma}^2_{zxx}\tilde{\sigma}^4_{zxy}\tilde{\sigma}^4_{zyy}\left(\tilde{\sigma}^2_{zxx}+2\tilde{\sigma}^2_{zyy}\right)
 \Bigg)
 \end{align}
is a monotonically decreasing function of $\tilde{z}$.  This can be verified by computing the
rate of loss of Fisher information as the beam propagates forward in the $z-$direction, given by $\partial_{\tilde{z}}{\bf{F}}_{bv}$. Again, the full expression for this quantity is quite long and uninformative, however, the important point is that this rate remains negative for all propagation distances, that is, the loss in Fisher Information is monotonic.  Plots of ${\bf{F}}_{bv}$, and $\partial_z{\bf{F}}_{bv}$ are presented and discussed below in Section~\ref{sec:Disc}.

\subsection{Plane Wave and Paraboloidal Beam}
Two commonly-discussed optical beams are the plane wave and the paraboloidal beam with intensities given by, respectively,
\begin{subequations}
\begin{align}
&{\text{Plane Wave}}\qquad\qquad\quad \rho(\tilde{x},\tilde{y},\tilde{z}|\tilde{\theta})=1,\\
&{\text{Paraboloidal Beam}}\qquad\rho(\tilde{x},\tilde{y},\tilde{z}|\tilde{\theta})=\dfrac{1}{k^2z^2}.
\end{align}
\end{subequations}
Each of these beams solves the PWE exactly but the solutions are trivial because there is no $(\tilde{x},\tilde{y})$-dependence to the transverse field distribution. As a result, the Hessian, the optical pressure and the Fisher information are all zero. Hence, as defined, these beam do not diffract but they are also non-integrable and thus not physically realizable.


\section{Discussion }\label{sec:Disc}

In Section~\ref{sec:minFIM} we showed that diffraction is a minimizer of the Fisher information associated with any parametric model of a beam's intensity. That is, optical intensity will move locally in such a manner as to monotonically minimize ${\bf{F}}$ -- a global parameter -- during propagation. However, there is an important caveat, namely, that the Hessian must be defined at the local point. For example, consider the case of the HG$_{10}$ beam in Section~\ref{sec:Examples} that exhibits a minimum at the origin. But the Hessian (\ref{eq:HessianHG01}) for this intensity distribution is undefined at the origin. Hence, in this case, no intensity flows in toward the origin as confirmed by the acceleration plots in Fig.~\ref{fig:HGDensityAccel}.

\begin{figure}[ht]
 \centerline{
  \begin{tabular}{c}
  \includegraphics[width=11cm,angle=0]
  {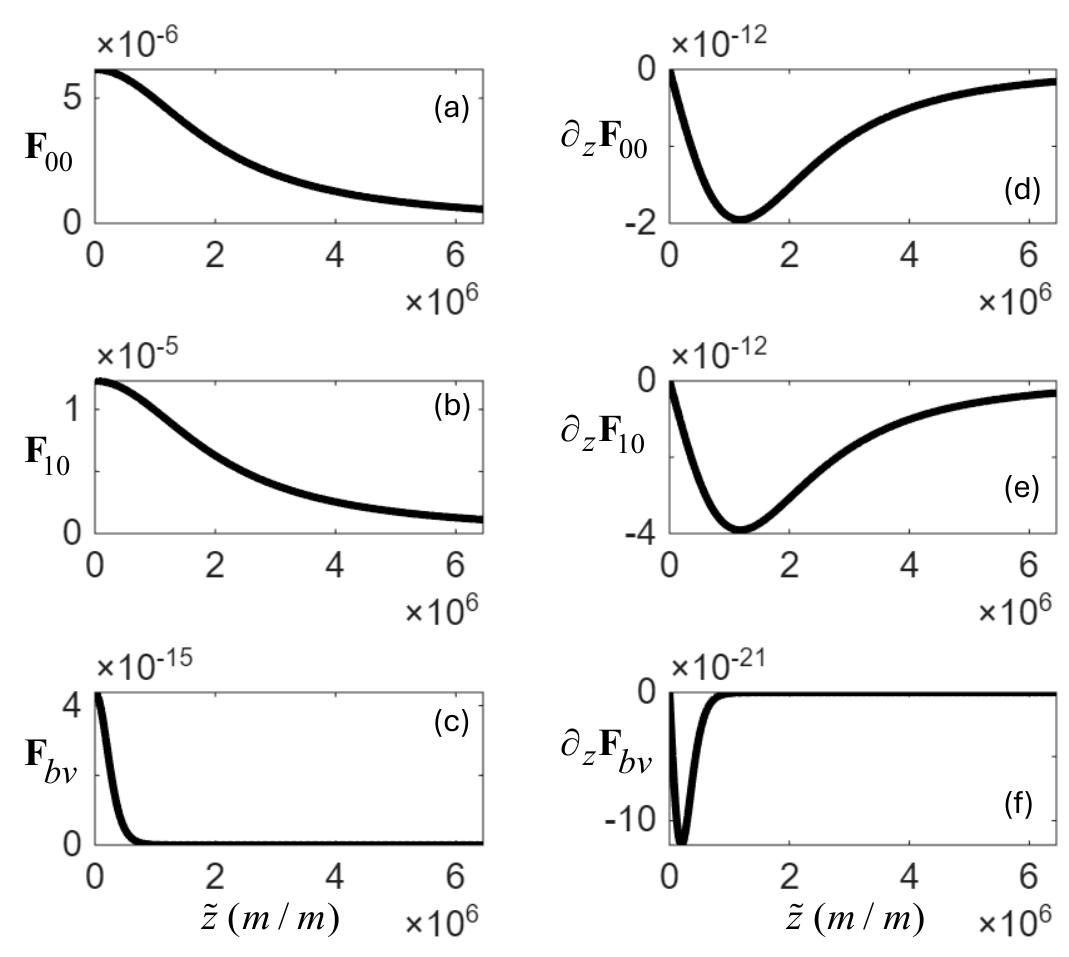}   
  \end{tabular}
  }
   \caption{(a), (b), (c): Plots of the Fisher information ${\bf{F}}$ as a function of $\tilde{z}$ for the three beam examples in Section~\ref{sec:Examples}. (d), (e), (f): Corresponding derivatives $\partial_z{\bf{F}}$ as a function of $\tilde{z}$. For these plots $\lambda=1.55\mu m$, $\sigma_{0x}=1.25\,mm,\,\sigma_{0y}=3.75\,mm,\,r=0.8$.}
   \label{fig:FanddFPlotsSummary2}   
\end{figure}

Figure~\ref{fig:FanddFPlotsSummary2} summarizes the $\tilde{z}-$dependent behavior of ${\bf{F}}$ and $\partial{\bf{F}}$ for all three beams.
\\
From these curves we can make the following observations.
\begin{itemize}
\item All three ${\bf{F}}$ curves approach zero monotonically and asymptotically as $\tilde{z}$ increases -- consistent with the general behavior argued by Frieden in 1989 \cite{Frieden:89} and as we derived above in Section~\ref{sec:minFIM}.
\item At any $\tilde{z}-$value, the absolute value of the Fisher information, ranked from largest to smallest, is ${\bf{F}}_{10}\rightarrow {\bf{F}}_{00}\rightarrow {\bf{F}}_{bv}$ and where the information content of the bivariate Gaussian beam is almost nine orders of magnitude smaller than that of the other two beams. Since the probability density associated with each intensity distribution was normalized to the value at the origin, these relative values of ${\bf{F}}$ must represent \emph{intrinsic} information content corresponding to the \emph{shape} of each distribution.
\item The bivariate Gaussian beam depletes its Fisher information at a significantly significantly faster than the other two beams.
\end{itemize}

It is also clear from Fig. (\ref{fig:FanddFPlotsSummary2}) that the absolute value of FI is closely tied to the number of model parameters used to describe that distribution.  This is a sensible result given that the FI quantifies the amount of information that an observation -- such as an observed intensity distribution --  carries about the model parameters.  Hence, in general the more complex the model, the lower the FI.  For example, the three-parameter BV model has an FI value that is orders of magnitude smaller that of the CG and HG$_{10}$ beams.  However, when comparing the results of different intensity models we can normalize by the initial FI, ${\bf F}(z=0)$ and focus instead on the rate at which this relative FI is changing on propagation.

\begin{figure}[tbh]
 \centerline{
  \begin{tabular}{c}
  \includegraphics[width=9cm,angle=0]
  {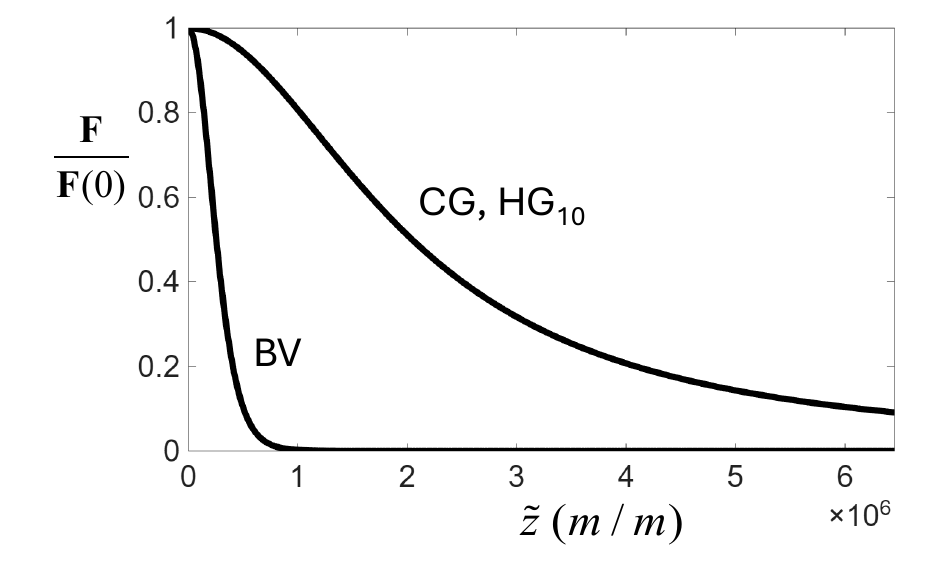}   
  \end{tabular}
  }
   \caption{Fisher information for each beam normalized to the value at $\tilde{z}=0$. BV = bivariate Gaussian, CG = circularly-symmetric Gaussian, HG$_{10}$ = \{1,0\}-Hermite Gaussian.}
   \label{fig:FnormSummary}   
\end{figure}

Figure~\ref{fig:FnormSummary} shows plots of Fisher information for each beam as a function of $\tilde{z}$ but now normalized to the initial value at $\tilde{z}=0$.
Comparing the normalized ${\bf{F}}$ curves we see that the CG and the HG$_{10}$  beams follow exactly the same curve, in spite of the fact that the distributions are quite distinct. This implies that, in these two examples, the dynamics producing information loss as the HG$_{10}$ beam propagates and diffracts are controlled mainly by the shape of the circularly-symmetric part of the beam and not by the higher-order spatial modulation $\mathcal{H}_{10}$ of the CG distribution $\rho_{00}$ (Eq.~\ref{eq:HGmode}). This effect is also evident in the plots of the acceleration streamlines in Figures \ref{fig:CGDensityAccel} and \ref{fig:HGDensityAccel}. It is seen that the directions of the acceleration vectors are nearly (but not exactly) identical in the two cases even though the Fisher information values, as seen in Figs.~\ref{fig:FanddFPlotsSummary2}(a) and \ref{fig:FanddFPlotsSummary2}(b), are different by a factor of two.

It is evident that the rate at which a beam loses Fisher information is directly tied to the rate at which the beam spreads (flattens) during diffraction.  This suggests, perhaps, a criteria for designing diffraction resistant beams via tailored phase profiles, for example, beams carrying orbital angular momentum \cite{Meng:21}.  The phase profile of such a beam possibly could be designed to minimize the loss of the Fisher information as a function of propagation distance $z$.  The resulting beam should suffer diffraction-induced spreading to a lesser extent than those with a faster decline in relative Fisher Information.  Although not yet tested,  this approach may represent a promising research direction based on the theory presented here.

We wish to emphasize a general point made by Frieden \cite{Frieden:04} regarding the important distinction between entropy and Fisher information in the context of continuous physical processes. Consider the discrete Shannon entropy
\begin{align}
H_X=\sum_{n=1}^N p_X(x_n)\log{(1/p_X(x_n))}.
\end{align}
Since the sum is taken over all possible outcomes $x_n$, $H_X$ is a global property of the distribution $p_X(x_n)$. Importantly, the sum is independent of the \emph{order} of the terms. Any rearrangement of terms will yield the same entropy and there is no relevance to a "local" ordering of terms. It is for this reason that the Shannon entropy and its thermodynamic counterpart, the Boltzmann entropy, find application in digital communications and statistical mechanics -- areas that are concerned with large collections of discrete, identical physical entities such as bits and particles. The Fisher information matrix likewise depends on a summation in the form of an integral but where the integrand contains the Hessian matrix which \emph{does} depend on the local arrangement of possible outcomes. Re-arranging the values $p_X(x)$ would completely change the shape of the distribution and hence the Hessian. Hence it is the Fisher information, and not the Shannon entropy, for example, that is more suitable for the description of continuous physical processes such as diffraction.

We make two concluding remarks.

Although not directly relevant for the discussion in this paper we note that the scalar diffraction theory is only valid in general when an optical beam is uniformly-polarized. In cases of non-uniform polarization distribution, such as a quadratic distribution of linear polarization \cite{Nichols:22, Nichols:24}, applying the scalar solution (\ref{eq:HFIntegral}) independently to the transverse vector electric field components can fail. By contrast, the transport model (\ref{eqn:transport}) and hence the results of this paper concerning the interpretation of diffraction as a minimizer of Fisher information remain valid.

Secondly, while we have chosen to draw our conclusions from the paraxial beam model, the interpretation of diffraction described herein extends to the more general case where this assumption is removed.  In this case, it can be shown \cite{Nichols:25} that the diffraction term has the exact same functional form, the only difference being the transverse gradient operator $\nabla_T$ is replaced by the full, three dimensional gradient operator.  Our subsequent interpretation of diffraction as an internal pressure is again seen to hold just as it did in the paraxial case.

\section{Summary}
Using the transport-of-intensity approach to beam propagation, we have shown that free-space optical diffraction is intimately connected to the Fisher information of the beam's intensity distribution, a connection that is hidden in the more-traditional scalar diffraction approach to beam propagation. We applied this approach to three example beams -- circularly-symmetric Gaussian, Hermite-Gaussian, and bivariate Gaussian -- and we calculated various beam propagation parameters including the Hessian and the  Fisher information matrix. We showed that the beams always propagate in a way such that the Fisher information continuously and monotonically approach zero although the rate at which the beam loses information depends on beam shape and position. There exist  applications such as free-space optical communications and directed energy where there is great interest in minimizing beam diffraction and the results developed here suggest minimizing the normalized Fisher information could be employed as a well-defined design criteria.

\vspace{6pt}

\bibliography{main}

@article{Meng:21,
author = {X. Meng and X. Chen and R. Chen and H. Li and T. Qu and A. Zhang},
journal = {Physical Review Applied},
volume = 16,
pages = 044063,
year = 2021,
title = {Generation of Multiple High-Order {B}essel Beams Carrying Different Orbital-Angular-Momentum Modes through an Anisotropic Holographic Impedance Metasurface}
}

@book{Price:06,
  title={Lagrangian and Eulerian representations of fluid flow: Kinematics and the equations of motion},
  author={Price, James F},
  year={2006},
  publisher={MIT OpenCourseWare}
}

@article{Ghosh:83,
author = {S. K. Ghosh},
title = {A Classical View of Quantum Chemistry},
journal = {Current Science},
volume = 52,
number = 16,
pages = {769-774},
year = 1983 
}

@article{Yahalom:18,
author = {A. Yahalom},
title = {The fluid dynamics of spin},
journal = {Molecular Physics},
year = 2018,
pages = {2698-2708},
volume = 116,
number = {19-20}
}

@book{Panton:96,
author = {R. L. Panton},
title = {Incompressible Flow},
year = 1996,
publisher = {Wiley}
}

@BOOK{Goodman:17,
      author       = {J. W. Goodman},
      title        = {{I}ntroduction to {F}ourier Optics; {F}ourth edition},
      address      = {New York},
      publisher    = {Macmillan Learning},
      year         = {2017}
}

@book{Whalen:95,
        Address = {San Diego},
        Author = {R. N. McDonough and A. D. Whalen},
        Edition = {Second},
        Publisher = {Academic Press},
        Title = {Detection of Signals in Noise},
        Year = 1995}

@article{Teague:82,
AUTHOR = {M.R. Teague},
JOURNAL = {Journal of the Optical Society of America},
TITLE = {Irradiance moments: Their propagation and use for unique retrieval of phase},
VOLUME = 72,
NUMBER = 6,
PAGES = {1199-1209},
YEAR = 1982
}

@article{Streibl:84,
AUTHOR = {N. Streibl},
JOURNAL = {Optics Communications},
TITLE = {Phase Imaging by the Transport Equation of Intensity},
VOLUME = 49,
NUMBER = 1,
PAGES = {6-10},
YEAR = 1984
}

@article{Zuo:20,
AUTHOR = {C. Zuo and J. Li and J. Sun and Y. Fan and J. Zhang and L. Lu and R. Zhang and B. Wang and L. Huang and Q. Chen},
JOURNAL = {Optics and Lasers in Engineering},
TITLE = {Transport of intensity equation: a tutorial},
VOLUME = 135,
PAGES = 106187,
YEAR = 2020
}

@article{Nichols:25,
title = {A General Model for Linearly Polarized Vector Beams},
author = {J. M. Nichols and F. Bucholtz},
year = 2025,
volume={2505.00471},
journal={arXiv},
}

@article{Nichols:24,
title = {Bending Light via Transverse Momentum Exchange: Theory and Experiment},
author = {J. M. Nichols and D. V. Nickel and F. Bucholtz},
year = 2025,
volume={112},
pages = 013517,
journal={Physical Review A},
}

@book{Frieden:04,
author = {Frieden, Roy},
year = {2004},
title = {Science from Fisher Information},
publisher = {Cambridge University Press}
}

@article{Frieden:89,
title = {Fisher Information as the Basis for Diffraction Optics},
author = {R. Frieden},
journal = {Optics Letters},
volume = 14,
number = 4,
year = 1989,
pages = {199-201}
}

@article{Nichols:22,
author = {J. M. Nichols and D. V. Nickel and F. Bucholtz},
title = {Vector beam bending via a polarization gradient},
journal = {Optics Express},
volume = 30,
number = 21,
pages = {38907-38929},
year = 2022
}

@article{Nichols:23,
author = {J. M. Nichols and D. V. Nickel and G. K. Rohde and F. Bucholtz},
title = {Transport model for the propagation of partially coherent, partially polarized, polarization-gradient vector beams},
journal = {Journal of the Optical Society of America - A},
volume = 40,
number = 6,
pages = {1084-1100},
year = 2023
}

@book{Saleh:91,
	Address = {New York},
	Author = {B. E. A. Saleh and M. C. Teich},
	Publisher = {John Wiley \& Sons, Inc.},
	Title = {Fundamentals of Photonics},
	Year = 1991}

\end{document}